\documentclass[a4paper]{article}

\usepackage{amsfonts,amsmath,amssymb,indentfirst,mathrsfs,amscd}
\usepackage[mathscr]{eucal}
\usepackage{amsthm}

\usepackage{graphicx}
\usepackage{hyperref}
\usepackage[nolist]{acronym}
\usepackage{subcaption}
\usepackage{multirow}
\usepackage{bm}

\usepackage{amsfonts}
\usepackage{authblk}

\usepackage{url}

\begin{acronym}
    \acro{WHO}{World Health Organization}
    \acro{IPVAW}{Intimate Partner Violence Against Women}
    \acro{DL}{Deep Learning}
    \acro{ML}{Machine Learning}
    \acro{AI}{Artificial Intelligence}
    \acro{RS}{Recommender System}
    \acro{ONS}{Office for National Statistics}
    \acro{CSEW}{Crime Survey for England and Wales}
    \acro{CF}{Collaborative Filtering}
    \acro{MF}{Matrix Factorization}
    \acro{MAE}{Mean Absolute Error}
    \acro{MSE}{Mean Squared Error}
    \acro{RMSE}{Root Mean Squared Error}
    \acro{CF4J}{Collaborative Filtering for Java}
    \acro{BeMF}{Bernoulli Matrix Factorization}
    \acro{RPI}{Reliability Prediction Index}
    \acro{KNN}{K Nearest Neighbours}
    \acro{PMF}{Probabilistic Matrix Factorization}
    \acro{NMF}{non-Negative Matrix Factorization}
    \acro{BNMF}{Bayesian non-Negative Matrix Factorization}
    \acro{URP}{User Ratings Profile Model}
    \acro{NN}{Neural Network}
    \acro{EU}{European Union}
    \acro{NC}{Nearest Centroid}
    \acro{LR}{Logistic Regression}
    \acro{KNN}{$K$-Nearest Neighbors}
    \acro{DT}{Decission Tree}
\end{acronym}

\newcommand{\cE}{\mathcal{E}}

\newcommand{\RR}{\mathbb{R}} 
\newcommand{\ZZ}{\mathbb{Z}} 

\begin{document}

\title{Machine learning for risk assessment in gender-based crime}

\author[1,2]{\'Angel Gonz\'alez-Prieto}
\author[3]{Antonio Br\'u}
\author[4]{Juan Carlos Nu\~no}
\author[5,6]{Jos\'e Luis Gonz\'alez-\'Alvarez}
\affil[1]{Departamento de Matem\'aticas, Universidad Aut\'onoma de Madrid, Madrid, Spain.}
\affil[2]{Instituto de Ciencias Matem\'aticas (CSIC-UAM-UC3M-UCM), Madrid, Spain.}
\affil[3]{Facultad de Ciencias Matemáticas, Universidad Complutense de Madrid, Madrid, Spain.}
\affil[4]{Departamento de Matem\'atica Aplicada, ETSI de Montes, Forestal y del Medio Natural,\newline{}Universidad Polit\'ecnica de Madrid, Madrid, Spain.}
\affil[5]{Gabinete de Coordinación y Estudios, Secretaría de Estado de Seguridad, Madrid, Spain.}
\affil[6]{Instituto de Ciencias Forenses y de la Seguridad (ICFS), Universidad Autónoma de Madrid, Madrid, Spain.}
\date{}                     
\setcounter{Maxaffil}{0}
\renewcommand\Affilfont{\itshape\small}


\markboth{Journal of \LaTeX\ Class Files,~Vol.~14, No.~8, August~2015}%
{Shell \MakeLowercase{\textit{et al.}}: Bare Demo of IEEEtran.cls for IEEE Journals}

\maketitle

\begin{abstract}
Gender-based crime is one of the most concerning scourges of contemporary society. Governments worldwide have invested lots of economic and human resources to radically eliminate this threat. Despite these efforts, providing accurate predictions of the risk that a victim of gender violence has of being attacked again is still a very hard open problem. The development of new methods for issuing accurate, fair and quick predictions would allow police forces to select the most appropriate measures to prevent recidivism. In this work, we propose to apply Machine Learning (ML) techniques to create models that accurately predict the recidivism risk of a gender-violence offender. The relevance of the contribution of this work is threefold: (i) the proposed ML method outperforms the preexisting risk assessment algorithm based on classical statistical techniques, (ii) the study has been conducted through an official specific-purpose database with more than 40,000 reports of gender violence, and (iii) two new quality measures are proposed for assessing the effective police protection that a model supplies and the overload in the invested resources that it generates.
Additionally, we propose a hybrid model that combines the statistical prediction methods with the ML method, permitting authorities to implement a smooth transition from the preexisting model to the ML-based model. This hybrid nature enables a decision-making process to optimally balance between the efficiency of the police system and aggressiveness of the protection measures taken.
\end{abstract}


\section{Introduction}\label{sec:introduction}

Despite the apparent regular occurrence of crime, as it was already recognized in the 19th century \cite{Durkheim,Quetelet}, it has defied the predictability provided by the scientific method in the Natural Sciences. Surprisingly, it is easier to accurately predict where a rocket will be after its launch in its way to a distant planet than to foresee the next victim of an offense. The unpredictable nature of crime arises the question of whether the classical scientific method can be a solving tool instead of only a descriptive framework. The Minority Report fiction of Philip K. Dick \cite{Philip} suggests that anticipating crime is possible and discusses some consequences derived from this possibility.  

In the real world, criminality can be handled by using mathematical and computational models \cite{Nunno,Shane}. Classical modeling has proven to be very accurate for some crimes with a noticeable spatial component, such as juvenile delinquency \cite{Brantingham,Short} or burglary occurrence \cite{Pitcher,Wang,Boque}. However, this kind of models seems to be less adequate to study other kinds of crimes that do not present a noticeable spatial component, as it is the case of domestic or gender-based violence. Instead, new computational methodologies under the name of \ac{ML}, a subfield of \ac{AI} \cite{Norvig}, appear as a powerful alternative \cite{berk2020algorithmic,Cabello,Hassan,Rodriguez2,Turner}. This approach focuses mainly on the development of computational algorithms to solve complex problems from the data \cite{Yang}. 

The consideration of gender-based crime as a singular case of offense is presented in many countries around the world. The USA legislation considers gender-based violence with its specificities but, nonetheless, a particularly focused tackling system does not exist. Along the last decades, several surveys have been released to get insight into the prevalence of different forms of violent crime in the USA and, in particular, they include information about the sex of the victims \cite{BJS}.

Across the Atlantic, most of the European countries have recognized the gender nature of domestic violence in the Istanbul Convention \cite{Istanbul}. 
The signatory governments of this protocol have subsequently adjusted their policies to fulfil the Convention’s requirements with adapted methods of fight and prevention.
In this sense, the UK deserves special mention since they did not ratify Convention's protocols. Instead, it has taken particular measures to fight against gender-based violence as a specific crime. Data obtained from the \ac{CSEW}, and published by the \ac{ONS}, concerning the gender component of violent crimes has been analyzed to answer the question whether gender-based crime is falling or not in the last times
\cite{Walby}. Additionally, the UK has implemented its own model for fighting gender crime, the so-called DASH (Domestic Abuse, Stalking and Honour Based Violence) system \cite{Dash}. The data gathered in this system has been analyzed in a recent paper using \ac{ML} techniques \cite{Turner}. The authors conclude that standard \ac{ML} models are not suitable for risk assessment. In this paper, we aim to defeat this claim. 


In particular, in this work we import cutting-edge computational methodologies of \ac{ML} to address the problem of prediction of recidivism in gender based-crimes. We shall focus on the particular case of \ac{IPVAW}, as defined by the \ac{WHO} \cite{world2013responding}. This definition excludes other offenses related with gender such as stalkings or rapes, which are crimes with a very different nature, more macroscopic and less victim-focused.

Given a large amount of structured data about \ac{IPVAW} cases, we will apply \ac{ML} techniques to develop novel models of risk assessment of recidivism of a victim, understood as the probability that a female victim, who has been offended and has reported her case, is aggressed again. In our case, the data will be provided by the Spanish VioGen system, a governmental program for tracking and controlling gender violence \cite{Belen,Torrecilla}, but the approach and applied methods are general and can be straightforwardly translated to other data sources.

Notice that providing highly accurate risk assessments is crucial for developing effective police protection. In practice, police resources are limited and, with the current investment in gender violence prosecution, it is not possible to provide close surveillance and active protection to all the victims. For this reason, an accurate prediction with high specificity is needed to categorize gender violence cases according to their risk during the first report. Through these predictions, it is possible to balance the resources, providing a level of protection according to the expected risk.

To empirically validate these \ac{ML}-based models, we conducted several experiments. As it will be evidenced by these results, the \ac{NC} model clearly outperforms the other risk-assessment methods. \ac{NC} is a powerful similarity-based classification method that has been successfully applied in cancer class prediction from gene expression profiling \cite{tibshirani2002diagnosis} and in text classification \cite{rocchio1971relevance}, where it is usually known as the Rocchio classifier. In our scenario, the \ac{NC} algorithm seeks to extract the main features of each of the archetypes of aggressors and, using them, analyzes new cases by computing the similarity to each of these general patterns. This procedure is analogue to some criminalistic methods \cite{vettor2013offender,del2020action}, but the large amount of data and the variety of responses allows the \ac{ML} to extract very subtle information which cannot be obtained via classical methods.

As baseline for these experiments, we will compare the \ac{ML}-based approach with the preexisting risk-assessment model. In our case study, this baseline model is implemented in VioGen. Based on classical statistics, this VioGen assessment model has been in production for more than ten years and has lead to an outstanding reduction in the reported gender crimes in Spain \cite{lopez2020intimate}. To this aim, we introduce two novel quality measures to evaluate the effectiveness of a prediction model: 1) provided police protection, which quantifies the extent to which the threaded victims are protected; and 2) police resources overload, which regards the amount of resources that are wasted due to an overcautious prediction of the model. A large value of the former metric points out to highly effective predictions and a low value of the later indicates an optimal use of the resources available.

The empirical results show that the \ac{ML}-based method clearly outperforms the preexisting model in terms of police protection and, depending on the features of the police system, it may also improve the usage of the resources or lead to a slight overrun. In order to migrate from the existing system to the \ac{ML}-based method, which may awaken initial reluctance, we also propose a hybrid model that allows decision-makers to implement a smooth progressive transition, avoiding a drastic change of model. In this spirit, the ultimate aim of this work is not to radically substitute the existing risk-prediction models, but to complement, extend and refine their predictions to detect new hidden recidivist cases.




We hope that this work will have a crucial impact in the reduction of recidivism in gender violence. Only through a hand-to-hand collaboration between human and computational forces the mankind will be able to eradicate this dangerous scourge. In our opinion, the proposed method meets all the requirements to make the first move towards this important goal.  

\section{Prediction models}\label{sec:recidivism-models}

The prediction of the risk of recidivism can be approached as a classification problem. To fix notation, let us briefly review the main ideas of a classification task. The aim of a classification problem is to understand a very complex phenomenon that assigns, to each individual, a certain \emph{label} or \emph{category}. More precisely, an individual is represented as a point $x \in \RR^d$ for some $d > 0$ so that each of the components of $x$ should be understood as interesting \emph{features} of the individual. On the other hand, there is a finite set $\Lambda$ of labels in such a way that each individual is assigned to an element of $\Lambda$. In other words, there is a function
$$
	f: \RR^d \to \Lambda
$$
such that, for each individual $x \in \RR^d$, $f(x) \in \Lambda$ is the category of $x$.

The key problem is that in general this labelling function $f$ is completely unknown or the phenomenon under study is so complex that the assignment $f$ is intrinsically fuzzy. The aim of a classification model is thus to provide a reasonable ansatz
$$
	\hat{f}: \RR^d \to \Lambda
$$
such that $\hat{f}$ is as similar to the real assignment $f$ as possible.

\subsection{Recidivism prediction}\label{sec:recidivism-prediction}

In the particular context of \ac{IPVAW}, we shall use the reported aggression cases as individuals. To be precise, suppose that our data source provides us, for each closed case of aggression, two features:
\begin{enumerate}
    \item The circumstances that characterize the violence case, understood as victim and aggressor profiles, socio-economical characteristics and potential aggravating factors. They will be encoded as vectors $x \in \RR^d$. 
    \item The number of subsequent aggressions that each victim suffered after the first report. 
\end{enumerate}

In our case, the circumstances of the crimes and the number of recidivisms were extracted from the VioGen database (for a more detailed description, please refer to Appendix \ref{sec:materials-and-methods}).

As target labels, we set three levels of \emph{recidivism risk}
$$
	\Lambda = \left\{\textrm{No}, \textrm{ Low}, \textrm{ High}\right\}.
$$
To each closed aggression case $x$, we assign a category of $\Lambda$ according to the number of entries subsequent episodes of violence. Concretely,
\begin{itemize}
	\item \textbf{$f(x)=$ No}: The victim suffered no further aggressions.
	\item \textbf{$f(x)=$ Low}: The victim suffered one or two more attacks.
	\item \textbf{$f(x)=$ High}: The victim suffered three of more offenses of violence after the first report.
\end{itemize}

The threshold of 3 cases to distinguish between `Low' risk cases and `High' risk cases may seem rather arbitrary. To clarify its dependency, we have conducted an analysis of sensitivity for this threshold. The results showed that setting the threshold to 4 or 5 further aggressions leads to small deviations with respect to the original limit of 3. Thereby, these results evidence that the upcoming discussion is robust under small variations of this threshold. 


In the following, we propose models $\hat{f}$ for automatic violence prediction so that, when a new case $x^{\textrm{new}}$ appear, the issued assessment $\hat{f}(x^{\textrm{new}}) \in \Lambda$ can be used as prediction in order to determine the measures of police protection to be taken. To evaluate these models, we will assume that the data source also provides a model of preexisting risk assessment (typically, of statistical nature). In the particular case of VioGen, it implements a risk assessment model based on psychometric criteria (see Appendix \ref{sec:gender-based-dataset} for a more detailed description). This preexisting risk model will be used as baseline for the experimental setting.

\subsection{Supervised learning models}\label{sec:supervised-learning-models}

The proposal of this work is to construct the prediction models for recidivism prediction through \ac{ML}. These models are designed to automatically distil knowledge from the previously reported data. For the convenience of the reader, we sketch the main features of the \ac{ML}-based models.

A \ac{ML} model is fed with a finite set $D = \left\{x_1, \ldots, x_N\right\}$, usually referred to as the \emph{dataset}, as well as their real categories $y_i = f(x_i)$ for $1 \leq i \leq N$. In our setting, recall that the instances $x_i$ will be the `circumstances' that characterize the violence case (data 1.\ of the previous section) and $y_i \in \Lambda$ is the true recidivism risk computed from the number of subsequent aggressions (labeled data 2.\ of the previous section).

Using this dataset, the system seeks to a function $\hat{f}: \RR^d \to \Lambda$ such that it minimizes an error function $\cE(f)$. Several proposals exists in the literature but typically this error function takes the form
$$
	\cE(f) = \frac{1}{N}\sum_{i= 1}^N \delta(\hat{f}(x_i), y_i).
$$
Here, $\delta: \Lambda \times \Lambda \to [0, \infty)$ is a pre-fixed distance on the set of labels. If $\delta$ is taken to be the discrete metric ($\delta(y, y) = 0$ and $\delta(y, y')=1$ if $y \neq y'$), then $\cE(f)$ is the ratio of misclassified instances.


Particularly, if we focus our attention on a parametrized family of functions $f_{\theta}$ for parameters $\theta \in \RR^m$, the error function becomes a function of $\theta$,
$$
	\cE(\theta) = \cE(f_\theta): \RR^m \to \RR.
$$

In this context, the search of the best model reduces to an optimization problem on $\theta$, a process called the \emph{training} in the \ac{ML} jargon.
Several optimization algorithms may be applied for minimizing $\cE$, like gradient descent, linear/quadratic programming methods or genetic programming approaches.


Typically, the chosen optimization method requires to fix, previously to the training, some values that determine the concrete implementation of the optimization procedure (e.g.\ the step in the gradient descent method). These values, and any other parameter that must be fixed beforehand, are called the \emph{hyper-parameters}, since they are at a higher level than the system parameters: they tune the model and the method applied to seek the best parameters. No general-purpose hyper-parameter optimization method is known so experiments must be carried out by conducting an exhaustive search for the optimal hyper-parameters.

In this work, we have considered as potential models several standard \ac{ML} classification models, namely, \ac{LR}, \ac{DT}, \ac{KNN} and \acf{NC}. A fine hyper-parameter search has been conduced to select the best setting for each algorithm. Among them, the results clearly show that \ac{NC} provides the most accurate prediction. Further details of this thorough model selection process can be found in Appendix \ref{sec:model-selection}.

\begin{figure*}[!h]
	\begin{center}
	\includegraphics[scale=0.14]{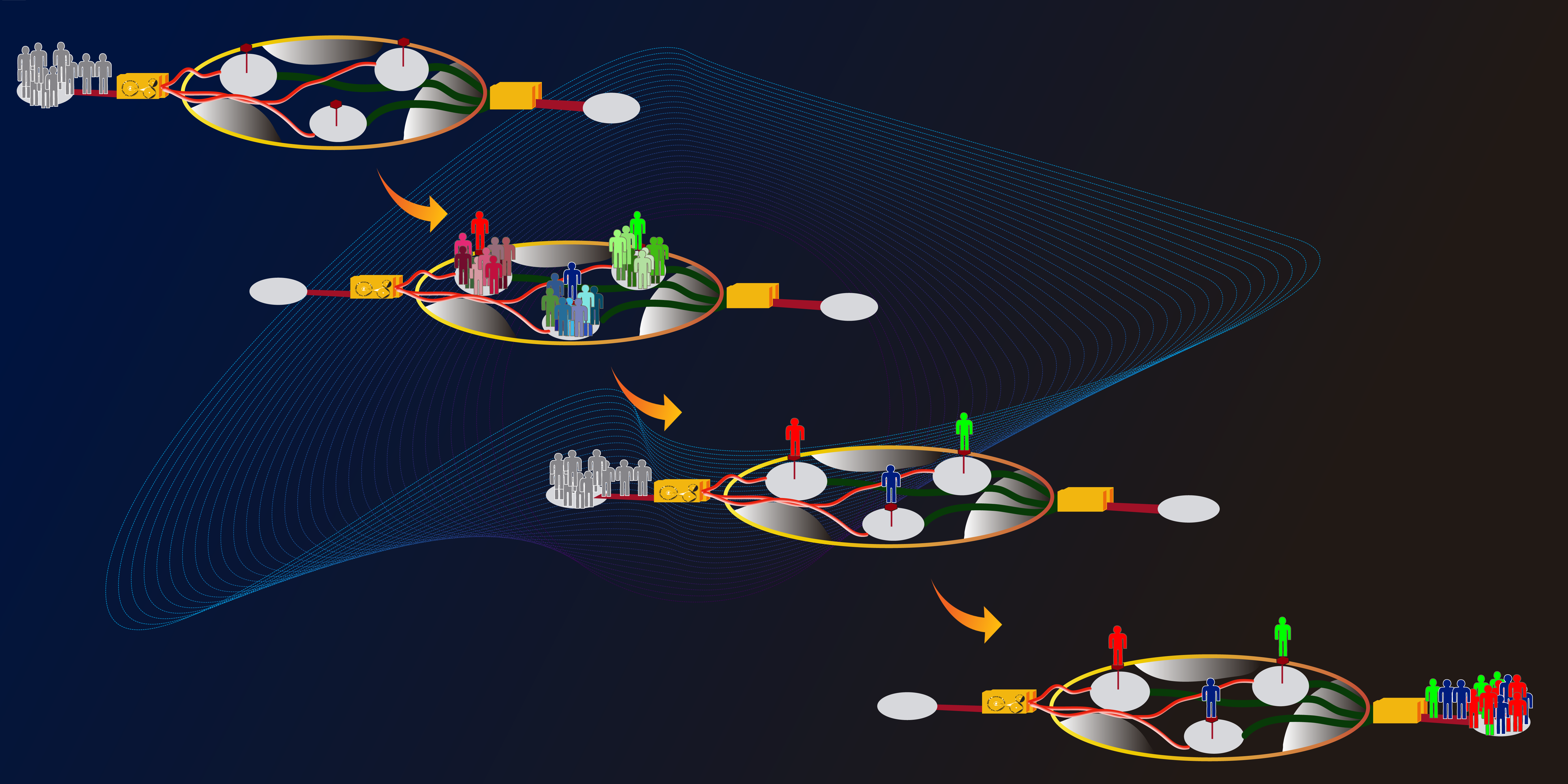}
	\end{center}
	\caption{Abstract representation of the \acf{NC} prediction method. Each level in the plot corresponds to a phase of the process. From top to bottom: initialization, training, system ready and exploitation.}
	\label{fig:model}
\vspace{-0cm}
\end{figure*}

A rough idea of the operation procedure of the \acf{NC} algorithm is shown in Figure \ref{fig:model}. During the initialization phase, shown in the upper-most picture, the model is only set up with the number of categories into which the cases must be classified (three in this case) but no further information is provided. In the training step, shown in the second picture of Figure \ref{fig:model}, the system is fed with a large number of examples for which the intended classification is known (or it may be calculable from other sources). In our case, the model analyzes the reports collected in the database and, according to the reported number of subsequent aggressions, classifies each case into `No' recidivism, `Low' risk of recidivism or `High' risk. Using these data, the system computes archetypal profiles of aggressions taking into account the provided features of the cases regarding the characteristics of the aggressor and of the victim, as well as the surrounding circumstances and aggravating factors. In \ac{NC} these profiles are extracted by computing the center-of-mass (the centroids) of each of the clusters according to some metric, typically the $L^p$-distance for some $p \geq 1$. An iterative shrinking process can also be applied so that the features which are close to the global centroid of the dataset are pulled apart. In this sense, the \ac{NC} training also includes a feature selection procedure through which the most relevant factors for recidivism are identified and extracted.

After this training process, the system has identified the profiles of each type of case and is ready to be used for production, as shown in the third level of Figure \ref{fig:model}. During the exploitation phase, shown in the forth level of Figure \ref{fig:model}, the system is able to classify accurately new cases of aggressions, even though very limited information is known, according to the similarity of the features of the analyzed case to each of the archetypal profiles. As by-product of this analysis, the model issues a prediction of the forecasted risk of recidivism (`No', `Low' or `High', represented in this figure by red, green and blue). This risk assessment can be used by the police forces to gauge the preventive measures to be provided to the victim in order to avoid further offenses.

\section{Experimental results}\label{sec:experimental-result}

In this section, we show the results of a collection of experiments conducted in order to analyze the performance of the proposed \ac{ML} models compared with existing risk-assessment method. 

\subsection{Police protection}\label{sec:police-protection}

In order to analyze the performance of the proposed approach against the rule system-based previous risk prediction model, we will consider an auxiliary quality measure.

Two quality measures are standard for classification problems, the so-called precision and recall. Let us fix a classification model $\hat{f}: \RR^d \to \Lambda$. Given a test set $x_1, \ldots, x_M$, let us denote the output of the classification model as $\hat{f}(x_1), \ldots, \hat{f}(x_N)$ whereas the real labels will be denoted by $y_1, \ldots, y_M$. Fixed a class $\lambda \in \Lambda$, the \emph{precision} and \emph{recall} of $\hat{f}$ in the class $\lambda$ are
\begin{align*}
	\textrm{Precision}_{\lambda}(\hat{f}) &= \frac{|\left\{1 \leq i \leq M\,|\, \hat{f}(x_i) = \lambda \textrm{ and } {y}_i = \lambda\right\}|}{|\left\{1 \leq i \leq M\,|\,\hat{f}(x_i) = \lambda\right\}|},\\
	\textrm{Recall}_{\lambda}(\hat{f}) &= \frac{|\left\{1 \leq i \leq M\,|\, \hat{f}(x_i) = \lambda \textrm{ and } {y}_i = \lambda\right\}|}{|\left\{1 \leq i \leq M\,|\,{y}_i = \lambda\right\}|},
\end{align*}
where $|X|$ stands for the number of elements of the set $X$. In other words, $1-\textrm{Precision}_{\lambda}(\hat{f})$ is the rate of false positives and $1-\textrm{Recall}_{\lambda}(\hat{f})$ is the rate of false negatives of the class $\lambda$.  In general, to combine both coefficients, it is customary to consider the \emph{$F_1$-score} as the harmonic mean
$$
	F_1\textrm{-score}_{\lambda}(\hat{f}) = 2 \frac{\textrm{Precision}_{\lambda}(\hat{f})\cdot \textrm{Recall}_{\lambda}(\hat{f})}{\textrm{Precision}_{\lambda}(\hat{f})+\textrm{Recall}_{\lambda}(\hat{f})}.
$$

Despite its wide range of application, precision and recall are not fair quality measures in our case. A prediction model with a good precision in the recidivism risk class 'High' but a bad recall might achieve an admissible $F_1$-score, but this is a very worthless model from the police viewpoint: it leaves most of the worst gender violence cases without police supervision. In the same vein, a bad precision in the 'No' class but a high recall is pointless, since the system is assigning no risk to very problematic cases.

In order to balance these opposed trends, we propose to use a novel measure of \emph{police protection}. Given a model $\hat{f}$, it is defined as
\begin{align*}
	\textrm{PoliceProtection}(\hat{f}) =& \textrm{Precision}_{\textrm{`No'}}(\hat{f}) + F_1\textrm{-score}_{\textrm{`Low'}}(\hat{f}) \\&+ \textrm{Recall}_{\textrm{`High'}}(\hat{f}).    
\end{align*}

In this way, the higher the police protection, the better the model. Algorithms achieving a large police protection are able to identify very violent cases with a high risk of recidivism. In our case, with appropriate hyper-parameter tuning, the \ac{NC} method reaches a police protection of $1.59$, in comparison with the preexisting (VioGen) risk prediction that can only achieve a police protection of $1.18$. For details on the conducted hyper-parameter optimization and the comparison with other algorithms, please refer to Appendix \ref{sec:model-selection}. Therefore, \ac{NC} clearly outperforms the police protection provided by previously implemented assessment methods.

\subsection{The hybrid model}\label{sec:hybrid-model}

Even though \ac{NC} reaches very good results in terms of police protection, it seems reasonable that the police forces would be reticent to drastically change the prediction method by a new untested one. Currently, VioGen is issuing fairly good predictions of risk and a drastic change could lead to a deterioration of the police efficacy if the model is not accurate in practice. In order to mitigate this reluctance, we propose to use a hybrid model between the preexisting risk assessment method and a \ac{ML}-based model.

The proposed model $\hat{f}^{\textrm{Hy}}_{\mu}$ is a stochastic mixture between the VioGen preexisting model and the \ac{NC} model. It depends on a real parameter $0 \leq \mu \leq 1$ in such a way that, for $\mu = 0$, the model $\hat{f}^{\textrm{Hy}}_{0}$ is the preexisting VioGen risk-assessment method and, for $\mu = 1$, the model $\hat{f}^{\textrm{Hy}}_{1}$ is the \ac{ML} algorithm. In the halfway, for an input $x \in \RR^d$, the output $\hat{f}^{\textrm{Hy}}_{\mu}(x)$ is a random variable that takes values in the oriented segment starting at the prediction $\hat{f}^{\textrm{Hy}}_{0}(x)$ and finishing at $\hat{f}^{\textrm{Hy}}_{1}(x)$ with mean the $\mu$ fraction of the segment. For a detailed description of the hybrid model, please refer to Appendix \ref{sec:hybrid-model}.

In order to test this model, we analyzed the dependence of the police protection metric, $\textrm{PoliceProtection}(\hat{f}^{\textrm{Hy}}_{\mu})$, with varying values of $\mu \in [0,1]$. For the purposes of these experiments, we used \ac{NC} with the previously chosen hyper-parameters as the \ac{ML} model. For the existing risk predictor, we took the cautious rule system for VioGen since it the setting that achieved a higher police protection score (see Appendix \ref{sec:gender-based-dataset} and Table \ref{tab:hyperparams-test} in Appendix \ref{sec:model-selection}). As shown in Figure \ref{fig:police-protection-hybrid}, an increase of the importance of the the \ac{ML} algorithm consistently leads to an increase of the police protection. Tests with other VioGen rule systems return similar results, and even show a more prevalent outperformance for the \ac{ML} method.

\begin{figure}[!h]
	\begin{center}
	\includegraphics[scale=0.75]{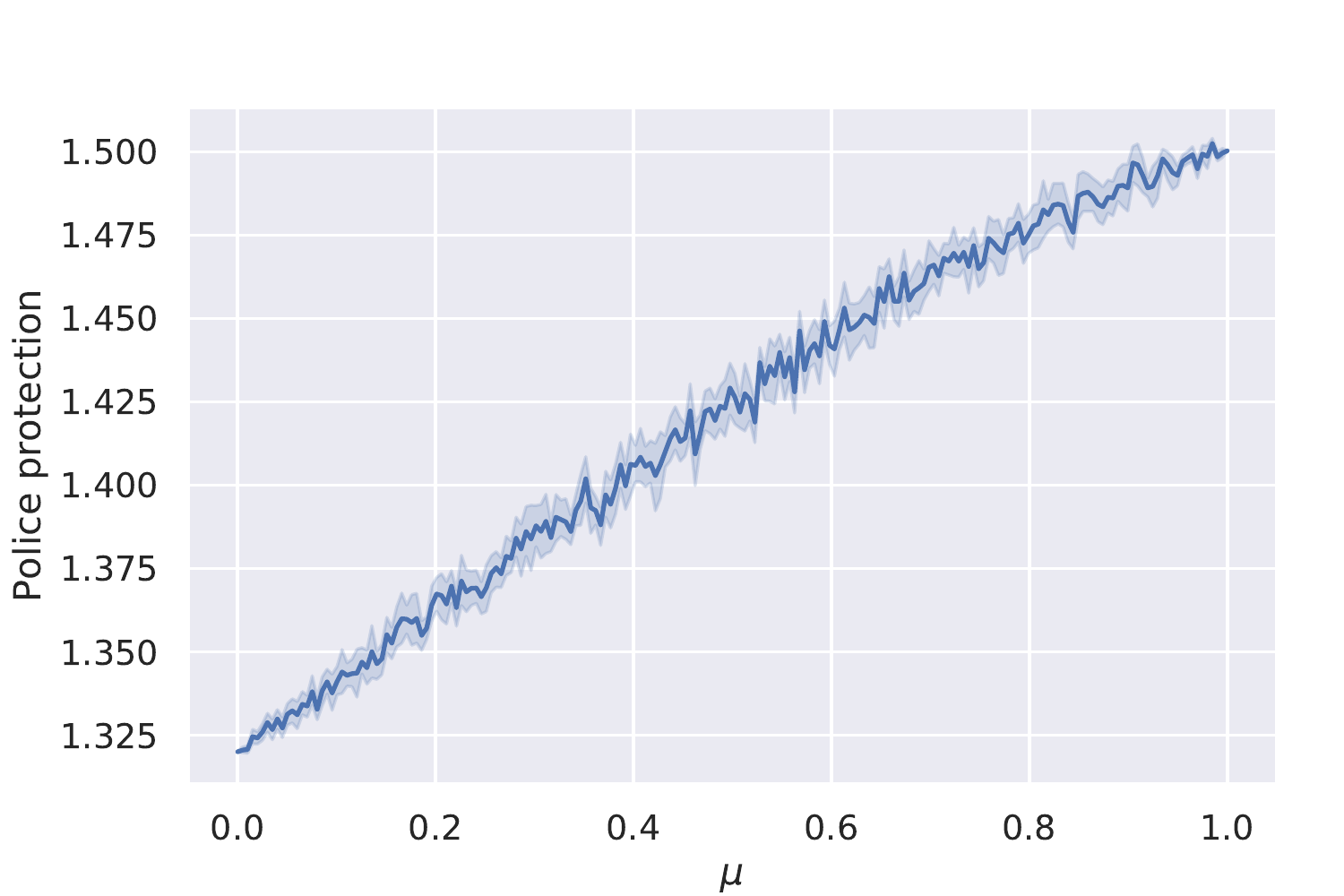}
	\end{center}
	\caption{Evolution of the police protection metric for the hybrid model with the best \ac{NC} model with varying values of $\mu$. For the hybrid predictor, the best \ac{NC} model (Metric = Euclidean and Shrink threshold = $0.1$) is used. The values of $\mu$ are taken from an uniform grid of the interval $[0,1]$ with $200$ equispaced points. For each value of $\mu$, a random sample of $10$ executions is considered. The solid line shows the mean value of the police protection metric along these executions, and the shadowed region is the $0.95$ confidence interval around this value. The length of this interval is short enough to provide sound evidence of behaviour of the evolution. Narrower confidence intervals are obtained for larger samples, with similar trends in the evolution of the metric.}
	\label{fig:police-protection-hybrid}
\vspace{-0cm}
\end{figure}

This idea can be implemented as a practical migration as follows. Initially, the prediction model used by the police forces might be this hybrid model with a small value of $\mu$ (say $\mu = 0.1$). For some time, the evolution of the recidivism statistics is screened. If, after a prudent time, the results improve the ones of the pure preexisting risk assessment system, the value of $\mu$ can be slightly increased and subsequently screened. In this way, a falling-off in the effectiveness of the system would be rapidly detected and actions may be taken to move back to the previous useful value of $\mu$. On the other hand, if the progression leads to a constant improvement, as predicted by Figure \ref{fig:police-protection-hybrid}, eventually the system would be completely migrated to \ac{ML} in a smooth and controlled way. 

\subsection{Resources overload}\label{sec:resources-overload}

An increase in the provided police protection might be in conflict with the amount of police resources mobilized. Indeed, there is an obvious solution with perfect police protection, namely to assign to all the cases the maximum risk. Of course, the problem is that this model is over-killed: it is mobilizing resources to provide protection to cases that many times will not need such a deployment. This is not only an economic issue. Since the resources are limited, this indiscriminate use of resources would lead to a degradation of the quality of the service, providing worse surveillance to the most threatened victims.

In order to quantify the unnecessary amount of police resources mobilized, we propose the following coefficient. Let $y_1, \ldots, y_M$ be the real values of the recidivism risk and suppose that $\hat{f}(x_1), \ldots, \hat{f}(x_M)$ are the issued predictions by a model $\hat{f}$. In addition, let us choose $\tau \geq 0$. Then, the \emph{police resources overload} coefficient of $\hat{f}$ with penalty $\tau$, denoted by $\textrm{PoliceResource}(\hat{f}; \tau)$, is just the weighted average of times $\hat{f}(x_i)$ is higher that $y_i$. Explicitly, it is given by
\begin{align*}
	\textrm{PoliceResource}(\hat{f};\tau) = & \frac{1}{2M(1+\tau)}\left(\left|\left\{\begin{matrix}\hat{f}(x_i) = \textrm{'Low'}\\ \textrm{ and } y_i=\textrm{'No'}\end{matrix}\right\}_i\right| + \tau\left|\left\{\begin{matrix}\hat{f}(x_i) = \textrm{'High'}\\ \textrm{ and } y_i=\textrm{'Low'}\end{matrix}\right\}_i\right| \right. \\
	&\,\,\,\,\,\,\,+ \left. (1+\tau)\left|\left\{\begin{matrix}\hat{f}(x_i) = \textrm{'High'}\\ \textrm{ and } y_i=\textrm{'No'}\end{matrix}\right\}_i\right|\right).
\end{align*}
The penalty $\tau$ should be understood as the extra overload that the police forces suffer when passing from a `Low' risk surveillance to a `High' risk surveillance. The particular value of $\tau$ depends on the policy of taken measures, as well as the particular structure, organization and role distribution of the police forces.

\begin{figure*}[!h]
\begin{subfigure}{0.53\textwidth}
	\begin{center}
	\includegraphics[scale=0.27]{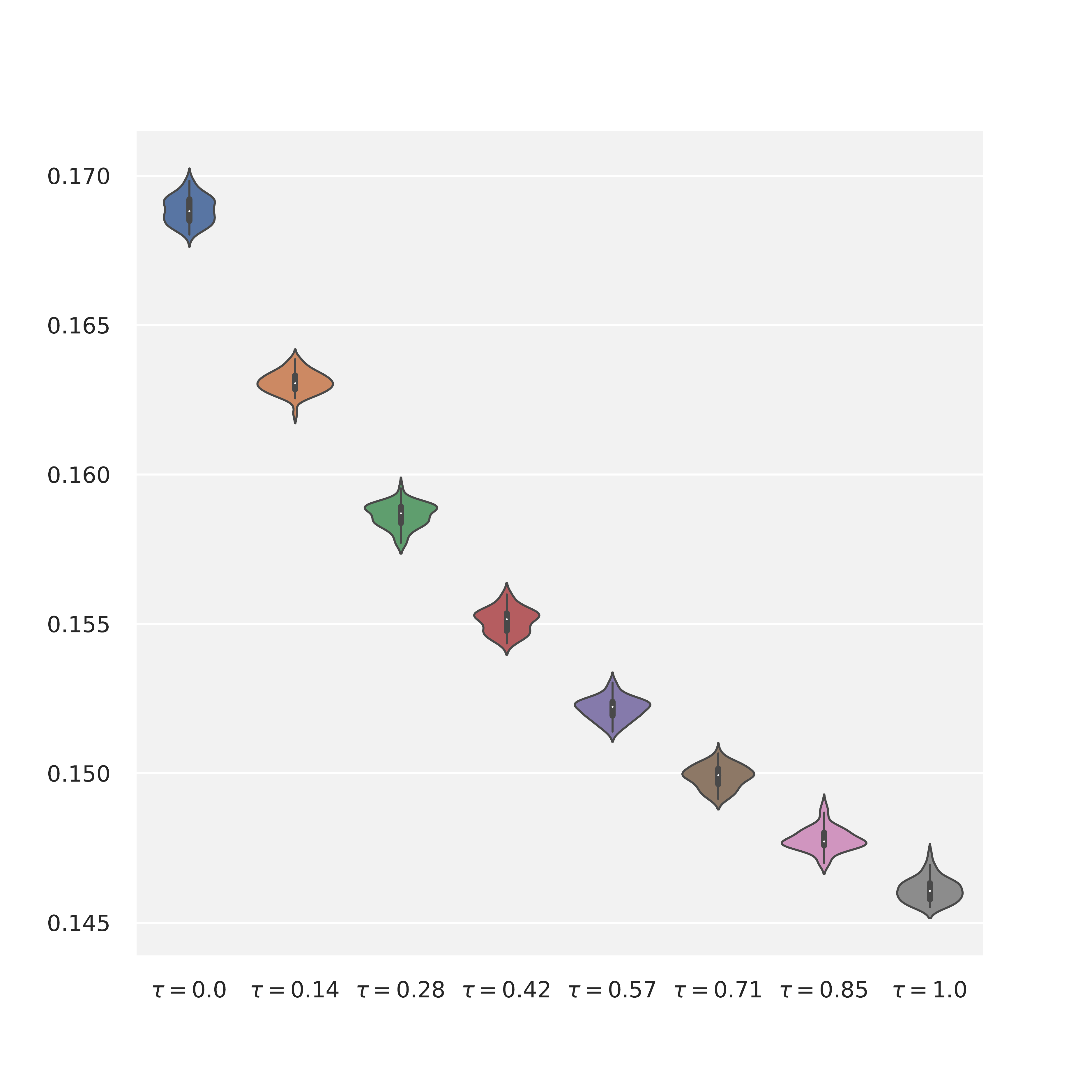}
	\end{center}
	\vspace{-0.9cm}
	\caption{Detail of the values of the police resources metrics for $\mu = 0.9$ for the hybrid model. Since the hybrid model is stochastic, a sample of $50$ executions was drawn. In horizontal, the results for different values of the penalty $\tau$. For each selected value of $\tau$, a boxplot of the main statistics of the obtained distribution for the police resources random variable is shown. Additionally, at both sides of these boxes we show the obtained empirical density function.}
	\label{fig:police-resources-violin}
\end{subfigure}\hspace{0.2cm}
\begin{subfigure}{0.46\textwidth}
    \begin{center}
	\vspace{2.75cm}
	\includegraphics[scale=0.57]{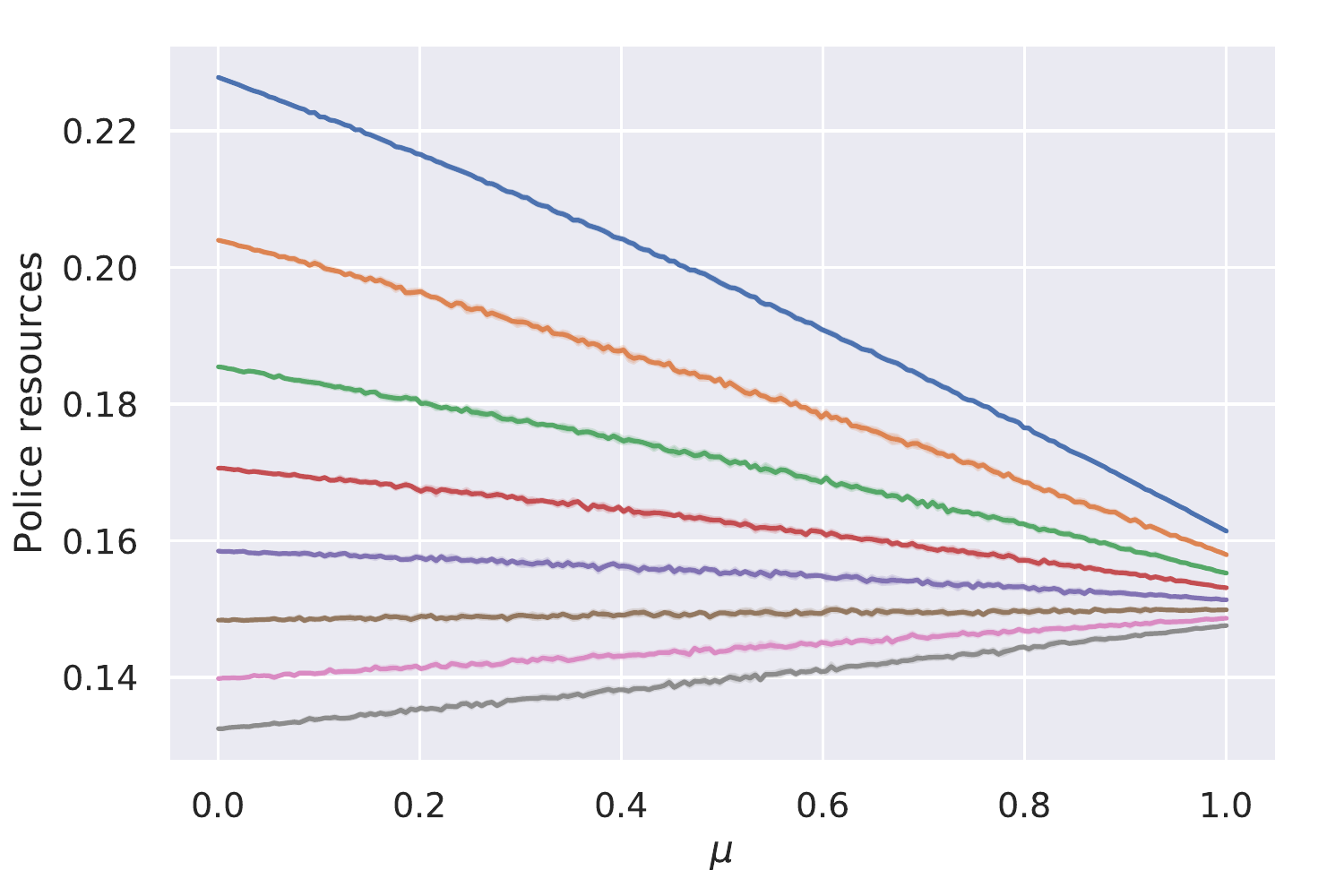}
    \end{center}
	\caption{Police resources metric for the hybrid approach for varying values of $\mu$. Each plot corresponds to the evolution of the police resources for a value of $\tau$, with the same code of colours as in Figure \ref{fig:police-resources-violin}. Each point was computed as the mean police resource overload obtained from a sample of $10$ executions. The mean amplitude of the $0.95$ confidence interval is $0.00064$ and the maximum amplitude observed is $0.00152$.}
	\label{fig:police-resources-hybrid}
\end{subfigure}
\caption{Results of the Police resources metric for different values of the penalty $\tau$ for the hybrid model. For the hybrid model, the best \ac{NC} model (Metric = Euclidean and Shrink threshold = $0.1$) is used. The parameter $\mu$ was uniformly sampled in the interval $[0,1]$ with $200$ sample points. }
\label{fig:police-resources}
\end{figure*}

In Figure \ref{fig:police-resources-violin} and \ref{fig:police-resources-hybrid} we show the police resources overload obtained on the test split of the dataset for the hybrid model $\hat{f} = \hat{f}^{\textrm{Hy}}_{\mu}$ with varying $\mu \in [0,1]$, as well as for various values of the penalty $\tau$. Despite the stochastic nature of the model, the results of the quality measure are consistent among executions. As it can be observed in Figure \ref{fig:police-resources-violin}, the obtained distributions exhibit a strong centrality with small standard deviations and $0.95$ confidence interval. The dynamics of the metric with varying $\mu$ for different values of the penalty $\tau$ is shown in Figure \ref{fig:police-resources-hybrid}. For small values of $\tau$, the overload of police resources is a decreasing function of $\mu$. In other words, a stronger weight of the \ac{ML} model in the hybrid model leads to a more efficient use of the resources for low penalties. When $\tau$ increases, the gain decreases and, around $\tau=0.5$, the trend inverts. For high penalties, the \ac{ML} model slightly increases the overload of the police system. Moreover, a clear trend is observed for all the values of penalties, with values between $0.15$ and $0.18$ for $\mu = 1$.

Summarizing, the results show that the proposed model based on \ac{NC} classification does not lead to a substantial increase in the invested police resources with respect to the existing prediction method, and even reduces it for small values of the penalty $\tau$. This points out that a progressive migration from the preexisting prediction model to a \ac{ML}-based model, like \ac{NC}, is highly encouraged.

\section{Discussion}\label{sec:discussion}

This work proposes a methodology based on \ac{ML} techniques to help authorities involved in policy against crime. In particular, it is applied to handle a official data set, VioGen, dependent of the Spanish Government, to assess the risk of revictimization in \ac{IPVAW}. The results obtained provide relevant clues to decision-makers to impose the right protection measures to victims. The ML algorithm is parametrically linked to the current VioGen system in order to take advantage of the previous experiences and, simultaneously, allows a successive improvement of risk assessment.

In this paper we have introduced a stochastic hybrid model that ensembles the prediction of the preexisting risk-assignment method with the proposed \ac{ML} method. This hybrid solution is parametrized by a coefficient $0 \leq \mu \leq 1$ that weighs the importance of the \ac{ML} model in the mixture. As aforementioned, this parameter $\mu$ allows decision-makers to smoothly transit from the existing risk prediction model to a fully \ac{ML} model. But, furthermore, this parameters also enables a fine tuning of the balance between invested resources and gain in the offered police protection. As shown in Figure \ref{fig:police-resources-hybrid}, for small values of the penalty $\tau$ the \ac{ML} method is more efficient than the existing risk-assessment model, so this transition towards a fully automatized model can be carried out without overloading the invested resources.

\begin{figure*}[!h]
    \begin{center}
	\includegraphics[scale=0.55]{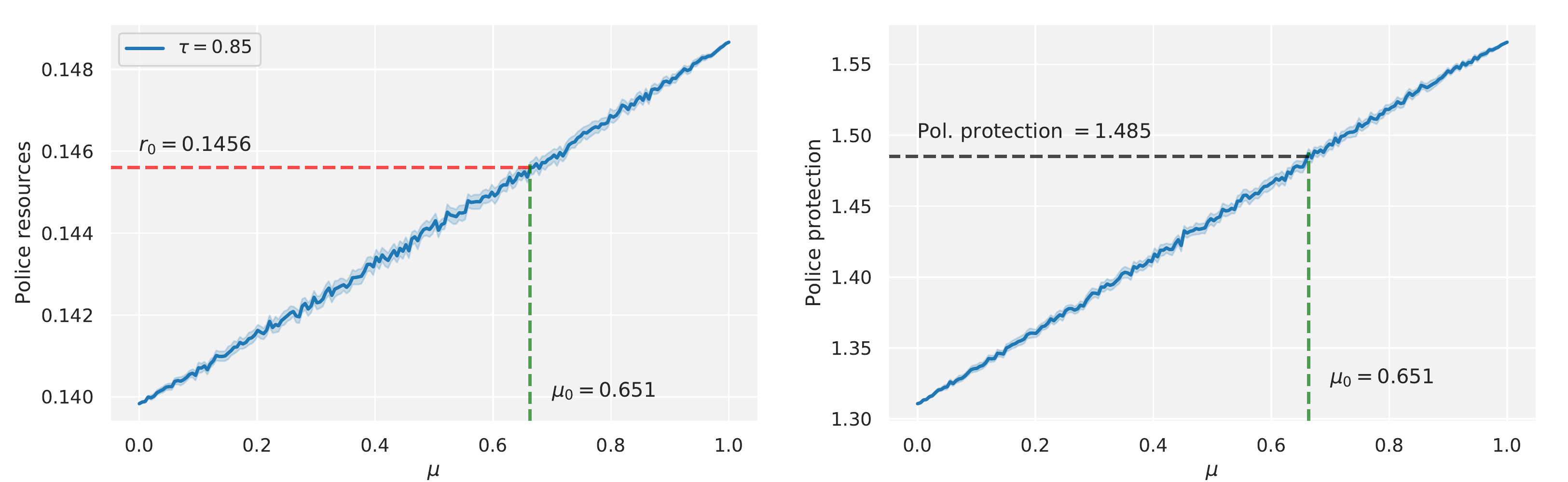}
    \end{center}
    \vspace{-0.3cm}
	\caption{Illustration of the procedure for adjusting the optimal value of $\mu$ with resource constraints. In this plot, the penalty of the police system has been set to $\tau = 0.85$ and the maximum acceptable resources overload to $r_0 = 0.1456$ ($4\%$ of increase with respect to the initial value of $0.14$). The line $y=r_0$ in the police resources plot (on the left) intersects the resources function at the optimal value $\mu_0$ ($\mu_0 = 0.651$ in this plot). The obtained gain in protection can be read from the police protection plot (on the right). In this example this value corresponds to $1.485$, which amounts to a increase of the $13\%$ with respect to the original value $1.31$. Therefore, an extra investment of the $4\%$ in resources leads to an improvement of the $13\%$ in the provided protection. For smaller values of $\tau$, the gain is even bigger.}
	\label{fig:police-decision}
\end{figure*}

Nevertheless, if the penalty $\tau$ is large, the \ac{ML} model suffers a small increase in the use of resources that might not be affordable for the police system. In that case, the two plots of Figures \ref{fig:police-protection-hybrid} and \ref{fig:police-resources-hybrid} may be jointly used by a decision-maker to provide an optimal solution. This process in illustrated in Figure \ref{fig:police-decision}. The first step is to empirically estimate the penalty $\tau$ of the police system, by conduction of in-field measures. Once $\tau$ has been calculated, the decision-maker should fix a maximum amount of resources that the system is able to support. Let us call this value $r_0$. Now, we draw the horizontal line $y=r_0$ in the police resources plot corresponding to $\tau$. The $x$-coordinate of the intersection of this line with the police resources function determines the maximum feasible value of the parameter $\mu$, namely $\mu_0$. For any value of the weight $0 \leq \mu \leq \mu_0$, the obtained hybrid system fulfils the imposed constraints in the used resources. Therefore, transitioning from the initial model $\mu = 0$ to the hybrid state $\mu = \mu_0$ leads to an increase in the provided police efficiency without exceeding the maximum of resources assigned.

Before ending, we would like to point out that the difficulties of risk prediction are not due to the dataset, which is certainly large and reliable, but to the intrinsic characteristics of the problem. The implementation of any of the measures that are
deduced from the system protocol is changing the own behavior of the
players in the crime scene. Unlike in other settings, such as medical research, it is not possible to set a control group where we do not intervene to detect how much this protocol improves safety. Due to obvious legal and moral concerns, it is not possible to choose not to act under the suspicious of danger. 

This lack of ground truth prevents any true validation of the model, leading to an intrinsic difficulty to measure whether the protective
actions have been really effective in the prevention of revictimization: if nothing wrong happens it may be a consequence of the properly taken actions or because the system failed in its prediction of danger. This cause-effect reversion is somehow analogous to the uncertainty principle of Quantum Mechanics, where measuring the system inevitably leads to its modification.

Besides, the dependent variable, the assessment of risk has never been measured. Instead, it has been only estimated from an indirect measure: the relapses. As a matter of fact, the algorithm is learning to classify from an intangible asset: the risk of victimization. As a consequence of these two conceptual drawbacks, standard \ac{ML} techniques are not always efficient, being \acf{NC}, as showed in this paper, an exception. Despite this apparent paradox, any policy program is going to be successful if crime rate is decreasing, independently of the cause.

In parallel with the \emph{Precog} system of Philip K. Dick, designed to eliminate criminality completely from our societies, the
algorithm of machine learning proposed in this paper uses the
information of previous violent episodes to provide an assessment of
risk for victimization. The results suggest that any security system, 
whatever its implementation within its national policy programs, would benefit from a rigorous risk classification following the main strategies of this work. Eventually, this analysis and application will be translated into a significative
drop in gender-based crimes. Furthermore, this approach also generalizes to other types of crimes, like juvenile delinquency and gangs rivalry, provided that some similarity assumptions hold, namely distinguishable profiles of aggressors and victims.

\appendix

\section{Gender-based crime dataset}\label{sec:gender-based-dataset}

The data processed in this paper was collected through the Spanish VioGen system. VioGen is a governmental computational system oriented to collect, analyze and provide preventive measures related with the cases of \ac{IPVAW}. Currently, it has a national scope of action and supports concurrent multiuser access. It provides support for archiving and parallel processing all the reports of gender crime that take place countrywide. According to these reports, the system issues a prediction of the recidivism risk for each case.

According to official sources, since its start up in 2007 until December 2020, 613,065 cases have been reported to the VioGen system  \cite{statsVioGen}. This implies that, on average, the system adds 75 new cases per day in a total population of 48 million people. On a regular basis, around 60,000 cases remain active at any given time, which require a huge amount of police resources. 

Among other functionalities, the VioGen system gathers data about the first occurrence of a \ac{IPVAW} case and its subsequent follow-ups. The former information is collected through a standard form called VPR, from the Spanish `Valoración Policial del Riesgo' or Police Assessment of the Risk. The later data are collected by means of a different form called VPER, from the Spanish `Valoración Policial de la Evoluci\'on del Riesgo' or Police Assessment of the Evolution of the Risk. In the following sections, we describe the scope and outcome of these forms, which have served as input for the training and testing of the \ac{ML} models.

\subsection{VPR: Initial aggression report}\label{sec:vpr} Once a woman proceeds to report the occurrence of a gender-based aggression to the police, the system offers the agent a detailed form with several questions regarding the circumstances, severity and potential aggravating factors surrounding the crime. In this way, after finishing the needed inquiries for gathering all the evidences, the police agent uploads these data into the VioGen system through the VPR form. 

According to the input data, the system provides a risk assessment of recidivism, that ranges from `Not appreciated', with value 0, to `Extreme', with value 4. It is worth mentioning that, in the analyzed version, this scale does not measure the risk of prospective lethal crime, but of recidivism in the violence behavior, regardless of its severity, intensity or time lapse between assaults. The amount of cases rated with each risk level certainly remains constant along time, with very small variations. Table \ref{tab:risk-distribution} shows the distribution of the risk assessments in the active cases reported in VioGen for the last five years as at 31st December.

\begin{table}[!h]
\vspace{0.2cm}
\begin{center}
\caption{Classification of active cases into the five risk groups in the
VioGen system along the last five years. Notice that the distributions
are stable over time.}
\label{tab:risk-distribution}
\begin{tabular}{|c||c|c|c|c|c|}
    \hline \textbf{Year} & \textbf{2016} & \textbf{2017} & \textbf{2018} & \textbf{2019} & \textbf{2020} \\\hline\hline
    Active cases & $52,635$ & $54,793$ & $58,498$ & $61,355$ & $63,656$ \\\hline
    Not appreciated & $56\%$ & $50\%$ & $43\%$ & $50\%$ & $49\%$ \\\hline
    Low & $36\%$ & $42\%$ & $46\%$ & $39\%$ & $41\%$ \\\hline
    Medium & $7\%$ & $8\%$ & $10\%$ & $10\%$& $10\%$ \\\hline
    High & $0.3\%$ & $0.4\%$ & $0.4\%$ & $0.7\%$ & $0.7\%$\\\hline
    Extreme & $0.02\%$ & $0.03\%$ & $0.04\%$ & $0.02\%$ & $0.01\%$\\\hline
\end{tabular}
\end{center}
\vspace{-0.5cm}
\end{table}

According to this forecast, the police agent in charge decides to apply some preventive measures, which may range from a periodic report of the evolution of the case to preventive imprisonment of the aggressor.

\subsection{VPER: Follow-up of the aggression case}\label{sec:vper} It may happen that, despite the taken measures, a subsequent aggression takes place. In that case, the new report is not added to the VioGen database through the VPR form but through a specialized form called VPER. Static indicators, like the previous police records of the aggressor, lose importance in the VPER form and new dynamical indicators regarding the evolution of the aggresivity are added. In addition, the VPER subsystem provides support for collecting the periodic reports that the victims communicate during the surveillance period, even thought no further offenses occurs. 

Notice that, by means of these VPER forms, it is possible to count the number of recidivisms of a closed case. This number correlates with the recidivism risk, the amount of threat of a victim of being offended again. This risk is the target variable that we will study in the upcoming sections. There, we will discuss how this recidivism risk can be predicted beforehand so that an appropriate police protection can be applied to the victim, which hopefully leads to an effective avoidance of new violent episodes.

\subsection{Rule systems derived from VioGen}\label{sec:rule-system} As mentioned above, the VioGen system automatically assigns a risk to each VPR case according to the provided answers to the VPR checklist, ranging from `Not appreciated' (0) to `Extreme' (4). This VioGen prediction is generated through classical statistical methods. Briefly, it operates as follows. The possible responses to each of the items of the questionnaires were assigned with a weight depending on their observed importance based on psychometric criteria. Hence, given a report of gender violence, its score of recidivism risk is computed as the weighted sum of all its responses according to the pre-fixed weights. To discretize this score, thresholds to each level were imposed in such a way that the number of cases in each of the five risk classes agrees with the empirical distribution of these values found in a pilot study, as classified by experts. The scoring weights have been updated throughout the different versions on VioGen (see \cite{lopez2020intimate} for the latest version). For accessibility and
completeness, this work applies the coefficients corresponding to the
VioGen prediction method described in \cite{lopez2017risk} (see also \cite{lopez-ossorio-thesis}).

Using this assessment of risk, there is a direct translation into a rule system for predicting the recidivism risk. This rule system maps the five VioGen classes into the three classes of $\Lambda = \left\{\textrm{No}, \textrm{ Low}, \textrm{ High}\right\}$. If we restrict ourselves to maps preserving the ordering of severity of the VioGen classes, a rule system is thus implemented just by setting up two thresholds that define the changes of class.
To cover all the possibilities, in this work we considered four rule systems, ranging from a \emph{lax system}, which tends to predict low risk of recidivism, to a \emph{cautious system}, which tries to provide a more aggressive police response by assigning higher levels of risk. The particular chosen thresholds for these rule systems can be found in Table \ref{tab:rule-systems}. These assignments agree with the last follow-up of the VioGen risk assessment \cite{lopez2019protocolo}.


\begin{table}
\begin{center}
\caption{Rule systems for the VioGen risk assessment considered in this work. Each VioGen class is mapped into a class of $\Lambda$. In bold, the assignments that vary depending on the imposed thresholds.}
\label{tab:rule-systems}
\begin{tabular}{|c|c|c|c|c|c|}
\hline &  \multicolumn{5}{c|}{\textbf{VioGen risk-assessment class}} \\
\hline \textbf{Rule System} &  Not appr. &  Low & Medium & High & Extreme \\\hline
\textbf{Lax} & No & \textbf{No} & Low & \textbf{Low} & High \\
\textbf{Medium lax} & No & \textbf{Low} & Low & \textbf{Low} & High \\
\textbf{Medium cautious} & No & \textbf{No} & Low & \textbf{High} & High \\
\textbf{Cautious} & No & \textbf{Low} & Low & \textbf{High} & High \\\hline
\end{tabular}
\end{center}
\vspace{-0.5cm}
\end{table}

Notice that, instead of the five risk-assessment classes provided by VioGen, for the risk classification problem we use only the three labels of $\Lambda$. From a methodological point of view, labeling the cases with only three classes allows us to distinguish more effectively between highly concerning cases and not so risky aggressors. With more classes, the boundaries become very fuzzy and some clusters contain only a few instances, in agreement with what happens when the threshold between the `Low' and `High' classes is increased. This collapsing of the five classes of VioGen into a three-levels scale is also typical in the literature, as in the works screening the performance of the VioGen risk assessment for recidivism avoidance \cite{lopez2019protocolo}. Additionally, we have conducted experiments using more prediction classes and the results obtained are similar to the ones exposed in this work with only three classes.

\begin{figure*}[!h]
	\begin{center}
	\includegraphics[scale=0.35]{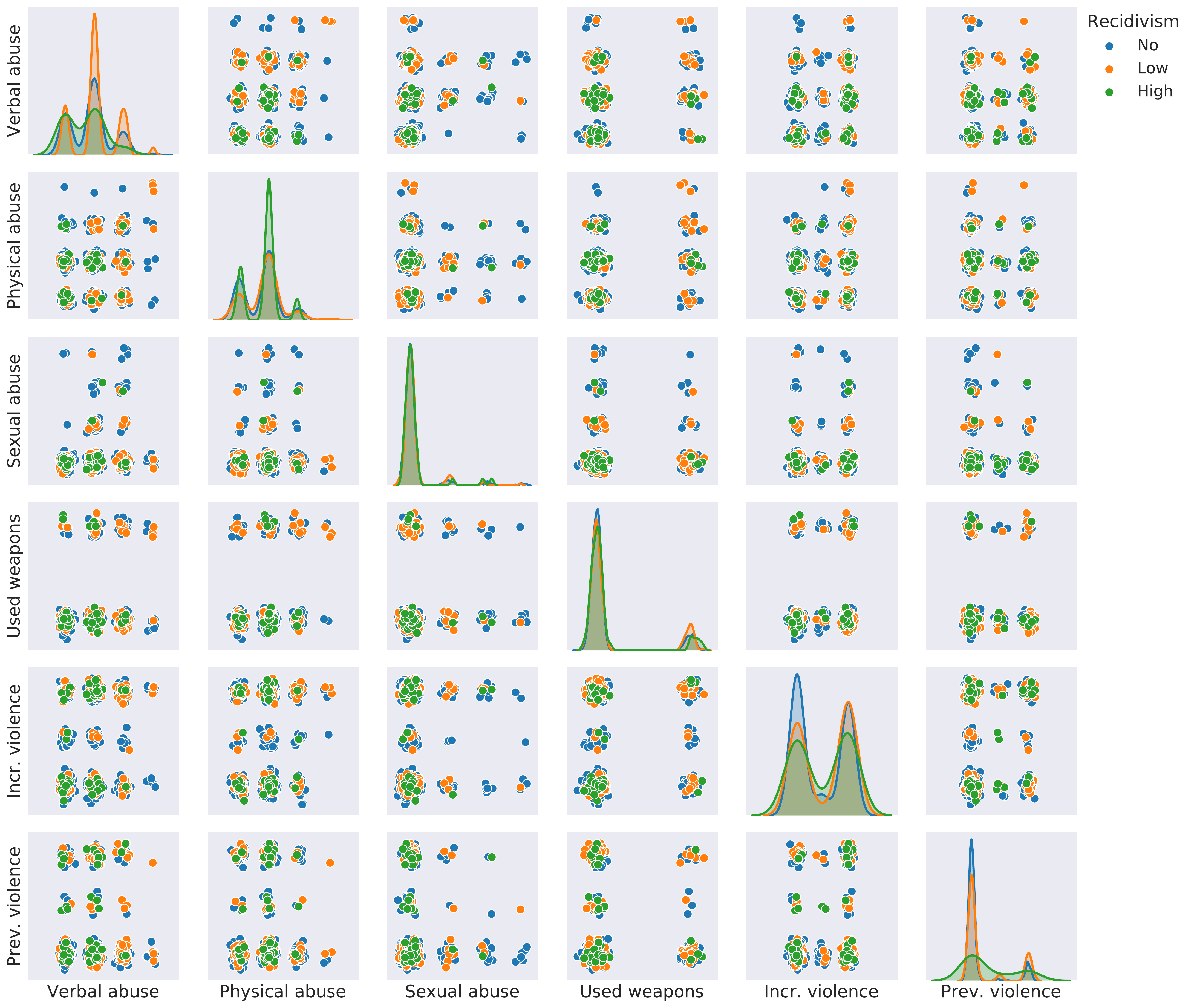}
	\end{center}
	\caption{Correlation plot of some of the collected answers to the VPR form. The plot compares the answers of $1000$ randomly chosen cases from the VioGen dataset. The diagonal plots show the distribution of the answers of the chosen questions. The off-diagonal plots compares two-by-two the answers to the questions. For each plot, the answers are placed on a rectangular grid with as many columns (resp.\ rows) as response options have the question displayed horizontally (resp.\ vertically). The left-most points of each plot for the horizontal axis (resp.\ bottom points for the vertical axis) correspond to `No'/`Very mild' responses, whereas the right-most points (resp.\ upper points) correspond to `Yes'/`Very severe' responses. Missing responses were assigned to a medium value. A small random noise was added to improve visualization.}
	\label{fig:correlation}
\end{figure*}

In spite of this reduction, the classification problem for risk assessment is still very hard. In Figure \ref{fig:correlation} we show the correlation matrix between some of the most relevant items in the VPR questionnaire with respect to each of the classes in $\Lambda$. Each point of the plot corresponds to a \ac{IPVAW} case and is coloured according to the observed recidivism with the codes explained in Section \ref{sec:recidivism-prediction}: Blue stands for `No' (no other offenses were reported), Orange stands for `Low' ($1$ or $2$ later offenses), and Green stands for `High' (more than $3$ subsequent offenses). The used indicators correspond to the following items of the VPR questionnaire: `Verbal abuse' refers to question F01-I01 (\textit{Did verbal abuse take place in the reported offense?}), `Physical abuse' refers to question F01-I02 (\textit{Did physical abuse take place in the reported offense?}), `Sexual abuse' refers to question F01-I03 (\textit{Did sexual abuse take place in the reported offense?}), `Used weapons' refers to question F02 (\textit{Did the aggressor use weapons or any other threatening objects against the victim in the reported offense?}), `Incr.\ violence' refers to question F04 (\textit{Does there exist an increasing severity and/or frequency in the aggressions or theats of violence in the last six months?}), and `Prev.\ violence' refers to question F08-I19 (\textit{Does there exist previous records of gender violence of the aggressor against other victims?}). This figure depicts the highly non-separable nature of the dataset, with many overlapping cases presenting similar responses to the main indicators.

Recall from Section \ref{sec:recidivism-prediction} that, to distinguish between `Low' and `High' risk cases, we set a threshold of $3$ recidivisms. This threshold may seem rather arbitrary but, for bigger thresholds $\geq 6$, the \ac{ML} models start losing accuracy in their predictions. The reason for this lost is twofold: first, the number of cases with $6$ or more offenses is very small ($\sim 100$ in our VioGen dataset) so the \ac{ML} method is not able to extract relevant features from such a limited dataset. Second, the selected \ac{NC} method operates by identifying `characteristic profiles' of \ac{IPVAW} cases. For high values of the threshold, the `Low' class aggregates several profiles of cases and this confuses the model, pointing out that a further refinement of the `Low' class is needed.

\section{Materials and Methods}\label{sec:materials-and-methods}

\subsection{Data preprocessing}\label{sec:data-preprocessing} The analyzed dataset is an excerpt of the Spanish VioGen database. The data correspond to 44,463 cases of \acf{IPVAW} reported between 1st October 2016 and 1st October 2017. Each case was stored as an array containing the 58 responses provided to the VPR form by the police agent when the aggression was reported. These questions collect information about the features of the violent episode, the psico-social features of the aggressor, the potential vulnerabilities of the victim, the circumstances related with minors and possible aggravating factors. A thorough description of these indicators and possible responses can be checked in \cite{lopez2017risk}. Demographic and temporal information, not corresponding to the VPR form but also collected in VioGen, was removed during the preprocessing phase to avoid bias. 

Missing values in the data were substituted by a special character. Multiple choice answers were codified through one-hot encoding, in such a way that each case is finally codified as a real $250$-dimensional vector that corresponds to an entry in the generated dataset. To compute the number of recidivist offenses, we counted the number of VPER forms associated to each case. Specifically, for each case number in the VPR database, associated to a case in the dataset, the number of VPER entries corresponding to aggressions with the same case number was assigned as the number of recidivisms.

\subsection{Reproducibility and data availability}\label{sec:reproducibility} The authors of this paper are committed to reproducible science. All the experiments have been conducted with the open source Python library  Scikit-learn library \cite{scikit-learn}. Due to privacy constraints, data are available under demand by contacting with the corresponding author.

\section{Model selection}\label{sec:model-selection}

The first step in the \ac{ML} pipeline is to conduct a hyper-parameter optimization on the models considered. Only after finding the best set of hyper-parameters, a fair comparison of different models can be carried out. To be precise, recall that the proposed model may depend on some hyper-parameters $\hat{f} = \hat{f}_\vartheta$. A standard criterion is to seek to the best setup $\vartheta_0$ such that $\vartheta_0 = \textrm{argmax}_\vartheta \left(F_1\textrm{-score}_{\lambda}(\hat{f}_\vartheta)\right)$. For that, a particular label $\lambda$ of interest may be fixed or the weighted average of the $F_1$-scores among all the labels may be consider.

In our case, we shall focus on the label $\lambda = $`High' and we will look for the hyper-parameters maximizing $\vartheta \mapsto F_1\textrm{-score}_{\textrm{`High'}}(\hat{f}_\vartheta)$. Four different types of models has been considered which are standard classifiers in \ac{ML}: Decision tree \cite{quinlan1986induction}, Random Forest \cite{ho1995random,breiman1996bagging}, K-Nearest Neighbors (KNN) \cite{altman1992introduction} and \acf{NC} \cite{tibshirani2002diagnosis}.

The dataset is randomly split into the training set (67\%) and the test set (33\%). For each model type and each possible combination of hyper-parameters, the model is trained with the training split and the quality score ($F_1$ for the class `High') is computed for the predicted values on the test split.

The search has been performed through a grid search with the following combinations of hyper-parameters: Decision tree (Criterion = Entropy, Gini; Splitter = Best, Random; Max depth = 5, 10, 50, 100, None), Random Forest (Criterion = Entropy, Gini; N estimators = 1, 5, 10, 100, 500; Max depth = 5, 10, 50, 100, None), K-Nearest Neighbors (K = 2, 5, 10, 20, 50, 100, 200) and Nearest Centroid (Metric = Euclidean, Minkowski, Manhattan; Shrink threshold = 0.1, 0.5, 1, 10, 20, None). For a precise description of the meaning of each of these hyper-parameters, please refer to the Scikit-learn library documentation \cite{scikit-learn}.

From these experiments, we obtain the optimal models of each type, which are shown in Table \ref{tab:hyperparams-test}. For each model we show the $F_1$-score attained in the class `High' (the objective function) as well as the weighted average of the $F_1$-score throughout all the three possible classes. The results evidence that the \ac{NC} model clearly outperforms the other models in the achieved $F_1$-score for the class `High'. The results for the weighted $F_1$ are similar in the four models (being slightly better in Decision trees). The scores obtained are consistent through the remaining sets of hyper-parameters. 

This superiority of the \ac{NC} model also holds in terms of police protection level, as can be checked in the fifth column of Table \ref{tab:hyperparams-test}. The police protection provided by the \ac{NC} algorithm clearly outperforms the ones obtained by the other analyzed methods, with an improvement around the $30\%$. In is worthy mentioning that this superiority is a post-hoc fact: the best method was not selected according to police protection metric but according to the $F_1$-score achieved in the `High' class, which is a standard criterion of model selection. A posteriori, we observe that this best method also outperforms regarding the police protection metric, even though it was not optimized for this task. This fact strengthens the reliability in the \ac{NC} method as the best model for addressing this problem. 

\begin{table}[t]
\begin{center}
\caption{Quality measures for the best prediction models against the test split. For each type of method, the best sets of hyper-parameters are shown ordered by rank according to their $F_1$-score for the `High' class (top is better). }
\label{tab:hyperparams-test}

\scriptsize
\begin{tabular}{|c|c|c|c|c|}
\hline \textbf{Method type} & \textbf{Hyper-parameters} & \textbf{\begin{tabular}{c}
    `High' class $F_1$  \\
    (higher is better) 
\end{tabular}} & \textbf{\begin{tabular}{c}
    Weighted $F_1$  \\
    (higher is better) 
\end{tabular}}  & \textbf{ \begin{tabular}{c}
    Police protection   \\
    (higher is better) 
\end{tabular}}\\\hline\hline
\multirow{7}{*}{\normalsize $\begin{matrix}\textrm{Decision}\\ \textrm{tree}\end{matrix}$} & $\begin{matrix} \textbf{\textrm{Criterion = Entropy}} \\ \textbf{\textrm{Splitter = Best}} \\ \textbf{\textrm{Max depth = None}} \end{matrix}$ &         \textbf{0.09} &        \textbf{0.62 } & \textbf{1.08}  \\\cline{2-5}
 & 
$\begin{matrix} \textrm{Criterion = Gini} \\ \textrm{Splitter = Random} \\ \textrm{Max depth = None} \end{matrix}$
 &         0.08  &      0.61       &  1.08  \\\cline{2-5}
 & $\begin{matrix} \textrm{Criterion = Entropy} \\ \textrm{Splitter = Random} \\ \textrm{Max depth = None} \end{matrix}$ &        0.08  &        0.61       & 1.08   \\\hline\hline
\multirow{7}{*}{\normalsize $\begin{matrix}\textrm{Random}\\ \textrm{forest}\end{matrix}$} & $\begin{matrix} \textbf{\textrm{Criterion = Entropy}} \\ \textbf{\textrm{N estimators = 1}} \\ \textbf{\textrm{Max depth = 50}} \end{matrix}$ &         \textbf{0.08} &        \textbf{0.62 } & \textbf{1.06} \\\cline{2-5}
 & 
$\begin{matrix} \textrm{Criterion = Entropy} \\ \textrm{N estimators = 1} \\ \textrm{Max depth = 100} \end{matrix}$
 &         0.08  &      0.61       &  1.05   \\\cline{2-5}
 & $\begin{matrix} \textrm{Criterion = Entropy} \\ \textrm{N estimators = 1} \\ \textrm{Max depth = None} \end{matrix}$ &        0.06  &        0.61       &   1.05  \\\hline\hline
\multirow{3}{*}{\normalsize $\begin{matrix}\textrm{K-Nearest}\\ \textrm{Neighbors}\end{matrix}$} & $\bm{K = 2}$ & \textbf{0.08} &        \textbf{ 0.57} &  \textbf{1.15}  \\\cline{2-5}
& $K = 5$ &  0.05   &  0.65      & 0.93  \\\cline{2-5}
& $K = 10$ &  0.00  &  0.65      & 0.84  \\\hline\hline
\multirow{5}{*}{\normalsize $\begin{matrix}\textrm{Nearest}\\ \textrm{Centroid}\end{matrix}$} & $\begin{matrix} \textbf{\textrm{Metric = Euclidean}}\\\textbf{\textrm{Shrink threshold = 0.1}} \end{matrix}$ &  \textbf{0.14}  &   \textbf{0.55}  &  \textbf{1.45}      \\\cline{2-5}
 & $\begin{matrix} \textrm{Metric = Euclidean}\\\textrm{Shrink threshold = 1} \end{matrix}$ & 0.14  &    0.54    &   1.48  \\\cline{2-5}
 & $\begin{matrix} \textrm{Metric = Manhattan}\\\textrm{Shrink threshold = 1}\end{matrix}$ & 0.13  &   0.57    &  1.35     \\\hline\hline
\multirow{4}{*}{\normalsize $\begin{matrix}\textrm{VioGen}\\ \textrm{risk}\end{matrix}$} & \textbf{Medium cautious} & \textbf{0.10} &        \textbf{0.62} &  \textbf{1.18} \\\cline{2-5}
 & Cautious & {0.10} &        {0.46} &  {1.28}  \\\cline{2-5}
 & Lax & {0.6} &        {0.63} &  {1.12} \\\cline{2-5}
 & Medium lax & {0.6} &        {0.47} &  {1.20}  \\\hline
\end{tabular}
\end{center}
\vspace{-0.5cm}
\end{table}

\subsection{Fine tuning for the \acf{NC} model} Once we selected \ac{NC} as the best \ac{ML} model for the risk prediction problem, we underwent a fine tuning of the hyper-parameters of the \ac{NC} model. To be precise, it was conducted a more exhaustive hyper-parameter search to optimize the police protection score with possible values `Metric' = Euclidean, Manhattan, Minkowski and `Shrink threshold' = 0.1, 0.25, 0.5, 0.75, 1, 5, 10, 20, None.
For this purpose, again the dataset was randomly split into the training set ($67\%$) and the test set ($33\%$). Now, on the training set, a cross validation procedure with $10$ folds is conducted with the police protection as objective function. The mean score and its standard deviation along the $10$ folds for the best setups of hyper-parameters are shown in Table \ref{tab:police-protection-cv}. According to these results, the best model is \ac{NC} with Euclidean metric and Shrink threshold = 5. This is the model that was used for the experiments of Section \ref{sec:experimental-result}.

Finally, notice that the rule system classifier obtained via the existing VioGen risk prediction achieved a police protection score of $1.28$ (see Table \ref{tab:hyperparams-test}).

\begin{table}
\begin{center}
\caption{Detailed police protection score of the \ac{NC} method in cross validation with $10$ folds.}
\label{tab:police-protection-cv}
\begin{tabular}{|c|c|c|c|}
\hline Parameters &  Mean &  Std & Rank\\\hline
 $\begin{matrix} \textbf{\textrm{Metric = Euclidean}}\\\textbf{\textrm{Shrink threshold = 5}} \end{matrix}$ &         \textbf{1.595} &        \textbf{0.037} &                \textbf{1} \\\hline
 $\begin{matrix} \textrm{Metric = Minkowski}\\\textrm{Shrink threshold = 5} \end{matrix}$ &         1.595 &        0.037 &                2 \\\hline
 $\begin{matrix} \textrm{Metric = Euclidean}\\\textrm{Shrink threshold = 1}\end{matrix}$ &     1.535 & 	0.045 &                3 \\\hline
 $\begin{matrix} \textrm{Metric = Minkowski}\\\textrm{Shrink threshold = 1}\end{matrix}$ &         1.535  &	0.045  &                4 \\\hline
\end{tabular}
\end{center}
\vspace{-0.5cm}
\end{table}

\section{Mathematical formulation of the hybrid model.}

Suppose that we have two prediction models with integral predictions $\hat{f}_0, \hat{f}_1: \RR^d \to \Lambda \subseteq \ZZ$. Additionally, suppose that $\Lambda$ is closed under convex combinations with integer coefficients. Given $0 \leq \mu \leq 1$ and $n \geq 0$, let $X_\mu(n) \in \textrm{Bin}(n, \mu)$ be a binomial random variable with success probability $\mu$ and $n$ trials i.e.\ $X_\mu(n) = \sum_{i=1}^n X_\mu^i$ where $X_\mu^i \sim \textrm{Ber}(\mu)$ are independent Bernoulli random variables with success probability $\mu$.

The hybrid model $\hat{f}^{\textrm{Hy}}_{\mu}$ assigns to an example $x \in \RR^d$ the value
$$
	\hat{f}^{\textrm{Hy}}_{\mu}(x) = \hat{f}_{0}(x) + \textrm{sign}(\rho)\, X_{\mu}(|\rho|),
$$
where $\rho = \hat{f}_{1}(x)-\hat{f}_{0}(x)$. In other words, $\hat{f}^{\textrm{Hy}}_{\mu}(x)$ starts with the initial prediction $\hat{f}_{0}(x)$. If $\hat{f}_{1}(x) > \hat{f}_{0}(x)$, the model tosses $|\hat{f}_{1}(x) > \hat{f}_{0}(x)|$ coins with success probability $\mu$ and, for each success, the system increases prediction by one unit. On the other hand, if $\hat{f}_{1}(x) < \hat{f}_{0}(x)$, the system reduces the prediction one unit per success. If the predictions $\hat{f}_0(x)$ and $\hat{f}_1(x)$ agree, the result remains unchanged.

Notice that, with this configuration, for $\mu = 0$ we have $\hat{f}^{\textrm{Hy}}_{0} = \hat{f}_{0}$, and for $\mu = 1$ we get $\hat{f}^{\textrm{Hy}}_{1} = \hat{f}_{1}$. Hence, $\hat{f}^{\textrm{Hy}}_{\mu}$ can be seen as a stochastic interpolation between the two models or a one-sided random walk with finite length. An analogous construction can be carried out if $\Lambda \subseteq \RR^n$ is a subset of a $m$-dimensional lattice, closed under convex combinations with integral coefficients, by running $m$ parallel independent hybrid models.

In the case considered in this paper, we have codified numerically the recidivism-risk labels in $ \Lambda$ ($0$ is `No' recidivism risk, $1$ is `Low' risk, and $2$ is `High' risk). The model $\hat{f}_{0}$ was taken as the numerical outputs of the preexisting VioGen risk assessment through a fixed rule system and $\hat{f}_{1}$ was the output of the considered \ac{ML} model.

\section*{Acknowledgment}
The authors are greatly indebted to D.\ G\'omez-Castro for reading very carefully this manuscript, his comments and suggestions to improve the exposition of this paper, and his constant encouragement throughout the development of this project. They also acknowledge the support of the staff of the Centro de Procesamiento de Datos (CPD) at Universidad Complutense de Madrid during the conduction of the experiments of this work. The first author also wants to thank R.\ Lara-Cabrera and F.\ Ortega for very useful conversations, as well as the hospitality of the ETSI de Sistemas Inform\'aticos at Universidad Polit\'ecnica de Madrid where this work was partially developed.

\bibliography{biblio}

\begin{thebibliography}{10}

\bibitem{altman1992introduction}
N.~S. Altman.
\newblock An introduction to kernel and nearest-neighbor nonparametric
  regression.
\newblock {\em The American Statistician}, 46(3):175--185, 1992.

\bibitem{berk2020algorithmic}
R.~A. Berk and S.~B. Sorenson.
\newblock Algorithmic approach to forecasting rare violent events: An
  illustration based in intimate partner violence perpetration.
\newblock {\em Criminology \& Public Policy}, 19(1):213--233, 2020.

\bibitem{Boque}
P.~Boqu{\'e}, L.~Serra, and M.~Saez.
\newblock `{S}urfing' burglaries with forced entry in {C}atalonia: Large-scale
  testing of near repeat victimization theory.
\newblock {\em European Journal of Criminology}, page 1477370820968102, 2020.

\bibitem{Brantingham}
P.~J. Brantingham, B.~Yuan, and D.~Herz.
\newblock Is gang violent crime more contagious than non-gang violent crime?
\newblock {\em Journal of Quantitative Criminology}, pages 1--25, 2020.

\bibitem{breiman1996bagging}
L.~Breiman.
\newblock Bagging predictors.
\newblock {\em Machine learning}, 24(2):123--140, 1996.

\bibitem{BJS}
{Bureau of Justice Statistics}.
\newblock Statistics about different types of crime in the {USA}.
\newblock \url{https://www.bjs.gov/index.cfm?ty=pbtp&tid=3}, 2021.
\newblock Accessed: 2021-02-05.

\bibitem{Cabello}
J.~G. Cabello.
\newblock Intimate partner violence: A novel warning system in which the
  victims' environment alerts to the danger.
\newblock {\em Heliyon}, 6(1):e03211, 2020.

\bibitem{Istanbul}
{Council of Europe}.
\newblock {\em The Council of Europe Convention on Preventing and Combating
  Violence against Women and Domestic Violence}.
\newblock 2014.

\bibitem{del2020action}
M.~del Mar Pecino-Latorre, J.~Santos-Hermoso, M.~del Carmen P{\'e}rez-Fuentes,
  R.~M. Patr{\'o}-Hern{\'a}ndez, and J.~L. Gonz{\'a}lez-{\'A}lvarez.
\newblock The action system model: a typology of {S}panish homicides.
\newblock {\em Frontiers in psychology}, 11, 2020.

\bibitem{Philip}
P.~K. Dick.
\newblock {\em The minority report: And other classic stories}, volume~30.
\newblock Citadel Press, 2002.

\bibitem{Durkheim}
E.~Durkheim.
\newblock {\em Le crime, ph{\'e}nom{\'e}ne normal}.
\newblock J.-M. Tremblay, 2006.

\bibitem{statsVioGen}
{Government Office against Gender-based Violence}.
\newblock Monthly statistical newsletter about gender-based crimes in {S}pain.
\newblock
  \url{https://violenciagenero.igualdad.gob.es/en/violenciaEnCifras/boletines/boletinMensual/home.htm},
  2021.
\newblock Accessed: 2021-02-05.

\bibitem{Hassan}
N.~Hassan, A.~Poudel, J.~Hale, C.~Hubacek, K.~T. Huq, S.~K.~K. Santu, and S.~I.
  Ahmed.
\newblock Towards automated sexual violence report tracking.
\newblock In {\em Proceedings of the International AAAI Conference on Web and
  Social Media}, volume~14, pages 250--259, 2020.

\bibitem{ho1995random}
T.~K. Ho.
\newblock Random decision forests.
\newblock In {\em Proceedings of 3rd international conference on document
  analysis and recognition}, volume~1, pages 278--282. IEEE, 1995.

\bibitem{Shane}
S.~D. Johnshon.
\newblock A brief history of the analysis of crime concentration.
\newblock {\em European Journal of Applied Mathematics}, 21(4-5):349, 2010.

\bibitem{lopez-ossorio-thesis}
J.~J. L{\'o}pez~Ossorio.
\newblock Construcci\'on y validaci\'on de los formularios de valoración
  policial del riesgo de reincidencia y violencia grave contra la pareja
  {(VPR4.0-VPER4.0)} del {M}inisterio del {I}nterior de {E}spaña.
\newblock 2017.
\newblock PhD Thesis. Facultad de Psicolog\'ia. Universidad Aut\'onoma de
  Madrid.

\bibitem{lopez2017risk}
J.~J. L{\'o}pez-Ossorio, J.~L.~G. {\'A}lvarez, S.~B. Pascual, L.~F.
  Garc{\'\i}a, and G.~Buela-Casal.
\newblock Risk factors related to intimate partner violence police recidivism
  in spain.
\newblock {\em International Journal of Clinical and Health Psychology},
  17(2):107--119, 2017.

\bibitem{lopez2020intimate}
J.~J. L{\'o}pez-Ossorio, J.~L. Gonz{\'a}lez-{\'A}lvarez, I.~Loinaz,
  A.~Mart{\'\i}nez-Mart{\'\i}nez, and D.~Pineda.
\newblock Intimate partner homicide risk assessment by police in spain: the
  dual protocol {VPR5.0-H}.
\newblock {\em Psychosocial Intervention}, 30(1):47--55, 2020.

\bibitem{lopez2019protocolo}
J.~J. L{\'o}pez-Ossorio, I.~Loinaz, and J.~L. Gonz{\'a}lez-{\'A}lvarez.
\newblock Protocolo para la valoraci{\'o}n policial del riesgo de violencia de
  g{\'e}nero (vpr4. 0): revisi{\'o}n de su funcionamiento.
\newblock {\em Revista Espa{\~n}ola de Medicina Legal}, 45(2):52--58, 2019.

\bibitem{Nunno}
J.~C. Nu\~{n}o, M.~A. Herrero, and M.~Primicerio.
\newblock A triangle model of criminality.
\newblock {\em Physica A: Statistical Mechanics and its Applications},
  387(12):2926--2936, 2008.

\bibitem{world2013responding}
W.~H. Organization.
\newblock {\em Responding to intimate partner violence and sexual violence
  against women: WHO clinical and policy guidelines}.
\newblock World Health Organization, 2013.

\bibitem{scikit-learn}
F.~Pedregosa, G.~Varoquaux, A.~Gramfort, V.~Michel, B.~Thirion, O.~Grisel,
  M.~Blondel, P.~Prettenhofer, R.~Weiss, V.~Dubourg, J.~Vanderplas, A.~Passos,
  D.~Cournapeau, M.~Brucher, M.~Perrot, and E.~Duchesnay.
\newblock Scikit-learn: Machine learning in {P}ython.
\newblock {\em Journal of Machine Learning Research}, 12:2825--2830, 2011.

\bibitem{Pitcher}
A.~B. Pitcher and S.~D. Johnson.
\newblock Exploring theories of victimization using a mathematical model of
  burglary.
\newblock {\em Journal of Research in Crime and Delinquency}, 48(1):83--109,
  2011.

\bibitem{Quetelet}
L.~A.~J. Quetelet.
\newblock {\em Recherches statistiques sur le royaume des Pays-Bas}.
\newblock Tarlier, 1829.

\bibitem{quinlan1986induction}
J.~R. Quinlan.
\newblock Induction of decision trees.
\newblock {\em Machine learning}, 1(1):81--106, 1986.

\bibitem{Dash}
L.~Richards.
\newblock Domestic abuse, stalking and harassment and honour based violence
  (dash, 2009) risk identification and assessment and management model.
\newblock {\em Association of Police Officers (ACPO)}, 2009.

\bibitem{rocchio1971relevance}
J.~Rocchio.
\newblock Relevance feedback in information retrieval.
\newblock {\em The Smart retrieval system-experiments in automatic document
  processing}, pages 313--323, 1971.

\bibitem{Rodriguez2}
I.~Rodr{\'\i}guez-Rodr{\'\i}guez, J.-V. Rodr{\'\i}guez, D.-J. Pardo-Quiles,
  P.~Heras-Gonz{\'a}lez, and I.~Chatzigiannakis.
\newblock Modeling and forecasting gender-based violence through machine
  learning techniques.
\newblock {\em Applied Sciences}, 10(22):8244, 2020.

\bibitem{Norvig}
S.~Russell and P.~Norvig.
\newblock {\em Artificial Intelligence: A Modern Approach}.
\newblock Prentice Hall Press, USA, 3rd edition, 2009.

\bibitem{Belen}
B.~Sanz-Barbero, C.~Linares, C.~Vives-Cases, J.~L. Gonz{\'a}lez, J.~J.
  L{\'o}pez-Ossorio, and J.~D{\'\i}az.
\newblock Intimate partner violence in madrid: a time series analysis
  (2008--2016).
\newblock {\em Annals of epidemiology}, 28(9):635--640, 2018.

\bibitem{Short}
M.~B. Short, P.~J. Brantingham, A.~L. Bertozzi, and G.~E. Tita.
\newblock Dissipation and displacement of hotspots in reaction-diffusion models
  of crime.
\newblock {\em Proceedings of the National Academy of Sciences},
  107(9):3961--3965, 2010.

\bibitem{tibshirani2002diagnosis}
R.~Tibshirani, T.~Hastie, B.~Narasimhan, and G.~Chu.
\newblock Diagnosis of multiple cancer types by shrunken centroids of gene
  expression.
\newblock {\em Proceedings of the National Academy of Sciences},
  99(10):6567--6572, 2002.

\bibitem{Torrecilla}
J.~L. Torrecilla, L.~Quijano-S{\'a}nchez, F.~Liberatore, J.~J.
  L{\'o}pez-Ossorio, and J.~L. Gonz{\'a}lez-{\'A}lvarez.
\newblock Evolution and study of a copycat effect in intimate partner
  homicides: A lesson from spanish femicides.
\newblock {\em PloS one}, 14(6):e0217914, 2019.

\bibitem{Turner}
E.~Turner, J.~Medina, and G.~Brown.
\newblock Dashing hopes? the predictive accuracy of domestic abuse risk
  assessment by police.
\newblock {\em The British Journal of Criminology}, 59(5):1013--1034, 2019.

\bibitem{vettor2013offender}
S.~Vettor, J.~Woodhams, and A.~R. Beech.
\newblock Offender profiling: A review and critique of the approaches and major
  assumptions.
\newblock {\em Journal of Current Issues in Crime, Law \& Law Enforcement},
  6(4), 2013.

\bibitem{Walby}
S.~Walby, J.~Towers, and B.~Francis.
\newblock Is violent crime increasing or decreasing? a new methodology to
  measure repeat attacks making visible the significance of gender and domestic
  relations.
\newblock {\em British Journal of Criminology}, 56(6):1203--1234, 2016.

\bibitem{Wang}
C.~Wang, Y.~Zhang, A.~L. Bertozzi, and M.~B. Short.
\newblock A stochastic-statistical residential burglary model with independent
  poisson clocks.
\newblock {\em European Journal of Applied Mathematics}, pages 1--27, 2020.

\bibitem{Yang}
Z.~R. Yang.
\newblock {\em Machine Learning Approaches to Bioinformatics}.
\newblock World Scientific Publishing Co., Inc., USA, 1st edition, 2010.

\end{thebibliography}

\bibliographystyle{abbrv}

\end{document}